\documentclass[sigconf, screen=true]{acmart}
\usepackage{enumitem}

\usepackage{todonotes}
\let\xtodo\todo
\renewcommand{\todo}[1]{\xtodo[inline,color=green!50]{#1}}

\AtBeginDocument{%
  }

\setcopyright{cc}
\copyrightyear{2026}
\acmYear{2026}
\acmDOI{https://doi.org/10.48550/arXiv.2603.21898}

\acmConference[]{Workshop on Data Literacy for the 21st Century at CHI 2026}{April 26}{Barcelona, Spain}

\begin{document}

\title{From Scores to Strategies: Towards Gaze-Informed Diagnostic Assessment for Visualization Literacy}


\author{Kathrin Schnizer}
\orcid{0009-0007-6952-2340}
\affiliation{%
	\institution{LMU Munich}
	\streetaddress{Frauenlobstr. 7a}
	\city{Munich}
	\country{Germany}}
\email{kathrin.schnizer@ifi.lmu.de}

\renewcommand{\shortauthors}{Schnizer}
\renewcommand{\shorttitle}{From Scores to Strategies}

\begin{abstract}
Visualization literacy assessments typically rely on correctness to classify performance, providing little evidence about how readers arrive at their answers. We argue that gaze can address this gap as an implicit process signal that complements standardized tests without sacrificing their scalability. Synthesizing findings from visualization and related research, we show that gaze metrics capture cognitive load invisible to accuracy and response time, and reflect strategy differences in attention allocation that track proficiency. We propose assessments that integrate literacy scores with gaze-derived process indicators — component-level attention profiles, integration frequency,
and viewing path dispersion — to distinguish fluent comprehension from labored success. This would shift literacy assessment from binary classification toward nuanced characterization of how readers navigate, integrate, and coordinate information across chart components. A roadmap identifies open challenges in empirical grounding, generalizability, assessment design, and practical feasibility.
\end{abstract}

\ccsdesc[500]{Human-centered computing~Information visualization}
\ccsdesc[500]{Human-centered computing~Empirical studies in HCI}

\keywords{Data Visualizations, Visualization Literacy, Eye-Tracking, Physiological Sensing}


\maketitle

\section{Introduction}

Visualizations are an effective means to communicate trends and patterns in data. As we generate and collect increasing amounts of data, it becomes crucial that people can correctly read and interpret data visualizations to extract meaningful insights, a skill often referred to as visualization literacy~\cite{lee2016vlat, boy2014principled, borner2016investigating}. To promote visualization literacy effectively, we need reliable ways to assess whether and how individuals understand a given visualization. Most visualization literacy assessments rely on explicit, outcome-based responses such as whether an answer is correct, rather than capturing how comprehension unfolds~\cite{nobre2024reading}. As a result, they primarily support classification of performance, offering limited evidence about attention allocation, reading strategies, and where processing difficulties
occur; information necessary for diagnosis beyond binary success or failure.

In practice, visualization literacy is often operationalized through performance metrics on standardized single-choice tasks~\cite{pandey2023mini, lee2016vlat, ge2023calvi}. These approaches are easy to administer and compare across studies, but they mainly reflect what users conclude, not how they navigated, prioritized, or integrated information to reach a conclusion. At the other end of the spectrum, free exploration and open-response approaches can reveal salient features, misconceptions, and reasoning~\cite{borner2019data, kim2020answering, rodrigues2021questions, quadri2024you}, yet they frequently require lengthy responses and intensive manual coding, increasing effort for both participants and evaluators. Together, these complementary approaches highlight the need for process-sensitive measures that can add diagnostic evidence to standardized assessments while remaining scalable.

In this position paper, we argue for augmenting outcome-based visualization literacy assessment with gaze as an implicit process signal. Our central claim is that gaze can bridge the two assessment needs identified above: the scalability of standardized tests and the diagnostic richness of qualitative approaches. We synthesize evidence
from visualization research showing that gaze metrics capture processing demands masked by correct responses and systematic strategy differences linked to proficiency. Building on this, we outline a vision of
assessments that report literacy scores alongside gaze-derived process indicators — such as attention distribution across chart components, integration between complementary elements, and viewing path dispersion — distinguishing fluent comprehension from labored success. We conclude with a roadmap addressing empirical grounding and diagnostic
interpretation, generalizability across items and readers, and the design of actionable assessment reports.

\section{Gaze for Process-Informed Visualization Literacy Assessment}
In the following, we motivate the use of gaze as process evidence for visualization literacy assessment. We first clarify what gaze can and cannot provide as a behavioral signal, then synthesize evidence from visualization research showing that gaze reflects processing demands that outcome measures miss and differentiates reading strategies linked to proficiency. Building on this evidence, we outline our vision for gaze-informed diagnostic assessment and a roadmap of open challenges.

\subsection{Gaze as a Process Trace of Attention Allocation}
\label{gaze-process-trace}
Eye-tracking metrics such as fixation durations, refixations, and transitions between regions provide time-resolved traces of where and when readers allocate overt visual attention while engaging with a data visualization. While gaze does not unambiguously reveal cognitive processing or intent, it provides direct behavioral evidence of attention allocation as it is shaped by task goals~\cite{oliveira2009discriminating, borji2015eyes}, bottom-up salience~\cite{itti2001computational}, and cognitive demands such as working memory load~\cite{brouwer2013distinguishing, walter2024quantifying}. Because gaze traces combine temporal and spatial information, they allow us to follow how attention is distributed across chart components over time (e.g., axes, marks, legend, or text), offering a window into reading and interaction processes that outcome measures alone cannot capture. 

\subsection{Gaze Reflects Processing Demands}
\label{sec:processing-demands}
Standardized visualization literacy tests such as the VLAT~\cite{lee2016vlat}, Mini-VLAT~\cite{pandey2023mini}, and CALVI~\cite{ge2023calvi} assess literacy through accuracy on task items. While this reveals which items a reader failed, it provides no signal about how demanding successfully completed items were for a given reader. Response time partially addresses this by capturing overall time on trials, and has been used to demonstrate that equivalent tasks vary in processing demands depending on visualization design~\cite{saket2018task, burch2011evaluation}. However, it does not reveal where in the chart effort was concentrated or which components impacted difficulty. As a result, a strong literacy score can mask processing struggles, limiting diagnosis to readers who fail rather than also identifying those whose correct responses mask effortful processing. 

Gaze metrics can address this gap by providing spatiotemporal evidence of how processing demands are distributed across chart components, even when the outcome is correct. The diagnostic value of such component-level process information has been demonstrated by~\citet{nobre2024reading}, who showed that capturing how readers engaged with chart elements during incorrect trials — through explicit input on reading strategies — enabled categorization of distinct literacy barriers beyond what accuracy alone could reveal. Gaze offers a pathway to obtain process evidence implicitly, extending the diagnostic potential to correct trials where struggles currently remain invisible.

At the level of individual gaze events, fixation frequency and duration, as well as saccade amplitude, have been shown to systematically vary with task complexity, with saccade size emerging as a particularly discriminatory parameter~\cite{chen2011eye}. Importantly, these metrics can be mapped to specific chart components when combined with area-of-interest
definitions. \citet{burch2011evaluation} demonstrated that readers produced more refixations on task-relevant elements in visualization layouts associated with longer completion times, indicating that spatial gaze patterns capture where processing effort concentrates at a granularity that response time alone cannot provide.

Complementing fixation-based metrics, pupil dilation provides a physiological index of cognitive load that is partially decoupled from overt viewing behavior. Task difficulty has been shown to systematically modulate pupil size~\cite{pomplun2019pupil}, a signal that has been leveraged to adapt system behavior in real time~\cite{kosch2018look, grootjen2024your}. In the context of visualization, \citet{toker2017leveraging} found that a visualization intervention which failed to improve task performance was associated with
larger pupil diameters, suggesting elevated cognitive load, demonstrating the kind of masked difficulty that accuracy-based assessment cannot detect. More broadly, \citet{van2023feedback} showed that eye-tracking signals explained approximately 50\% of variance in subjectively rated text comprehensibility, establishing that processing ease is recoverable from gaze as implicit feedback beyond accuracy.

It is worth noting that the evidence for masked difficulty currently derives from adjacent contexts: pupil dilation during visualization intervention evaluation~\cite{toker2017leveraging} and gaze-based comprehensibility detection during text reading~\cite{van2023feedback}. Neither directly demonstrates gaze detecting effortful-but-correct performance on a visualization literacy task. However, the convergence across modalities and task types supports the plausibility of this mechanism, and establishing it empirically within literacy assessment is a central objective of our roadmap outlined below.
Together, these findings support that if gaze reliably indexes processing difficulty beyond what accuracy and response time capture, it can surface cases where correct responses coincide with effortful processing. This would allow literacy assessment to distinguish fluent comprehension from labored success.

\subsection{Gaze Differentiates Reading Strategies Linked to Proficiency}
\label{sec:gaze-process-trace} 
Across studies that employ gaze to investigate visualization reading strategies, a consistent pattern emerges: proficient readers allocate attention in a more directed and integrative manner, while less proficient readers display more scattered and disconnected viewing behavior. Notably, this pattern holds across different operationalizations of proficiency: whether defined at the task level through correctness on individual items~\cite{netzel2017user, chang2025tell} or at the reader level through measured skill, expertise, or educational background~\cite{harsh2019seeing, toker2013individual, wu2025eye}. 
First, both successful and expert readers allocate more attention to contextual elements such as titles, legends, and axis labels, and do so in a directed rather than exploratory manner~\cite{harsh2019seeing, chang2025tell}. Second, proficient readers show more frequent transitions between chart components by moving between text and visual encodings~\cite{wu2025eye}, or between illustrations and accompanying text~\cite{jian2016fourth}, suggesting more active integration of complementary information sources. Third, refixations of previously attended regions have been linked to intermediate verification during successful task completions~\cite{netzel2017user}, pointing to a self-monitoring component in effective reading strategies. By contrast, less proficient readers tend toward scattered saccade patterns associated with inefficient search~\cite{wu2025eye, orlov2016eye} and show less spatial overlap in their attention across participants, suggesting more variable and less systematic approaches~\cite{chang2025tell}. These behavioral differences are not only tied to task outcomes but also track stable individual characteristics. Perceptual speed and verbal working memory have been associated with distinct gaze behaviors, including differences in transition patterns and the time allocated to legends and textual elements~\cite{toker2013individual}, while level of scientific education is correlated to the degree to which readers direct attention toward contextual chart elements~\cite{harsh2019seeing}.

The synthesized findings show that gaze captures coherent strategy differences in attention allocation and context integration, that systematically distinguish proficiency levels. This motivates complementing visualization literacy scores with gaze-derived strategy indicators that can pinpoint where proficient and struggling readers diverge.


\subsection{Vision: Gaze-Informed Diagnostic Visualization Literacy Assessment}
The evidence reviewed in the preceding sections establishes that gaze captures both processing demands (Section~\ref{sec:processing-demands}) and strategy differences linked to proficiency (Section~\ref{sec:gaze-process-trace}): information that accuracy and response time do not provide. We argue that this evidence motivates repurposing gaze from its current role in post-hoc evaluation of
visualization designs toward diagnostic assessment of individual readers. Concretely, we envision assessments that report a literacy score alongside gaze-derived process indicators: how attention is distributed across chart components, how frequently readers transition between complementary elements such as text and data encodings, and how concentrated versus dispersed their viewing path is. Such indicators would distinguish readers who arrive at correct answers through directed, integrative strategies from those whose success masks effortful or scattered processing. Retaining the standardized test format preserves administrative and scoring scalability; augmenting it with gaze adds a
diagnostic layer that can identify where and how readers struggle, even when their scores suggest otherwise.

\subsection{Proposed Roadmap and Challenges}
To move toward gaze-informed assessment, we outline four linked aspects that must be considered for gaze to function as diagnostic evidence alongside literacy scores:
\begin{enumerate}
    \item \textbf{Empirical Grounding and Diagnostic Interpretation:}
    Which gaze patterns, such as attention allocation across chart components, transitions between complementary elements, and refixation behavior, systematically co-vary with visualization literacy, and do they add explanatory value beyond correctness and response time?
    As discussed in Sections~\ref{sec:processing-demands} and ~\ref{sec:gaze-process-trace}, gaze traces are indirect measures of cognition that require interpretive grounding through triangulation
    with qualitative analyses of reading behavior. The open challenge is establishing this grounding systematically: linking observed gaze patterns to plausible sub-operations such as integration or
    verification in a way that supports diagnostic claims, not solely statistical associations.
    \item \textbf{Generalizability:} To serve as assessment evidence,
    gaze-derived indicators must be reliable under varying items, task types, and difficulty levels, and transferable across readers. While recent work on gaze prediction and simulation suggests learnable structure in viewing behavior on
    visualizations~\cite{shin2022scanner, wang2024visrecall,
    shi2025chartist, chang2025tell}, these efforts have largely focused on generating plausible traces associated with task success. For assessment, the open question is whether gaze strategies form stable, literacy-relevant signatures that generalize across items and readers, rather than reflecting item-specific or context-dependent variation.
    \item \textbf{Assessment Design:} How should gaze-derived process
    indicators be reported alongside literacy scores? This includes defining which indicators are presented (e.g., component-level attention profiles, integration frequency, path dispersion), how norms or reference distributions are established across proficiency levels, and whether indicators are reported per item, per reader, or both. Resolving these design questions determines whether gaze evidence remains a research tool or
    becomes a practical component of literacy assessment.
    \item \textbf{Practical Feasibility:} The preceding questions assume access to research-grade eye-tracking infrastructure, which is not universally available. For gaze-informed assessment to move beyond laboratory settings, it must be evaluated under realistic device and environmental constraints, including consumer-grade trackers with lower spatial and temporal resolution, variable lighting conditions, and uncalibrated setups. This raises the question of which gaze-derived indicators remain robust under degraded data quality, and whether the diagnostic distinctions established under controlled conditions transfer to less constrained settings.
\end{enumerate}
These questions frame the empirical, design, and practical challenges that must be addressed before gaze-informed diagnostics can complement standardized visualization literacy tests in practice.

\section{Reflection and Conclusion}
In this position paper, we argue that visualization literacy assessment should extend beyond outcome-based scoring to incorporate process evidence from gaze. The findings reviewed here establish that gaze captures both processing demands that accuracy and response time miss, and strategy differences in attention allocation and context integration that systematically track proficiency. Building on this evidence, we envision assessments that report literacy scores alongside gaze-derived process indicators, enabling a shift from classifying what readers get right or wrong to characterizing how they arrive at their answers to distinguish fluent comprehension from labored success, and directed strategies from scattered processing. 

Realizing this vision requires resolving the empirical and design challenges outlined in our roadmap. It also requires confronting the practical feasibility constraints outlined in our roadmap, particularly regarding deployment beyond controlled laboratory settings. More broadly, we hope this perspective contributes to ongoing discussions on how to operationalize visualization literacy and related constructs such
as data literacy. As gaze provides implicit, continuous evidence of how readers coordinate attention across information sources, the approach may extend to other data-centric tasks where comprehension depends on integrating multiple representations. We invite interdisciplinary discussion on what constitutes valid process evidence for such literacies, and what practical and ethical constraints shape its use.

\newpage

\section*{Transparency}
During the preparation of this work, the authors used Anthropic’s Claude for grammar and style editing. All content was reviewed and edited by the authors, who take full responsibility for the final publication.

\begin{acks}
Kathrin Schnizer was supported by the Deutsche Forschungsgemeinschaft (DFG, German Research Foundation) with Project ID 251654672 TRR 161.
\end{acks}

\bibliographystyle{ACM-Reference-Format}
\bibliography{bibliography}

\appendix









\end{document}